\documentclass[10pt,conference]{IEEEtran}

\usepackage{color}

\usepackage[pdftex]{graphicx}
\usepackage{amssymb}

\IEEEoverridecommandlockouts

\begin{document}

\title{FNT-based Reed-Solomon Erasure Codes\thanks{This work was supported by the French ANR grant No 2006 TCOM 019 (CAPRI-FEC project).}}

\author{
\IEEEauthorblockN{Alexandre Soro and J\'er\^ome Lacan}
\IEEEauthorblockA{University of Toulouse,\\ 
ISAE/DMIA, Toulouse, France\\ 
\{alexandre.soro,jerome.lacan\}@isae.fr}
}

\maketitle
%
%
%
%

\begin{abstract}
This paper presents a new construction of Maximum-Distance Separable (MDS) Reed-Solomon erasure codes based on Fermat Number Transform (FNT). Thanks to FNT, these codes support practical coding and decoding algorithms with complexity $O(n\log n)$, where $n$ is the number of symbols of a codeword. An open-source implementation shows that the encoding speed can reach 150Mbps for codes of length up to several 10,000s of symbols. These codes can be used as the basic component of the Information Dispersal Algorithm (IDA) system used in a several P2P systems. 
\end{abstract}

\section{Introduction and Related Work}
\label{introduction}
Erasure and network coding concepts were recently used in several kinds of networks in order to improve the throughput and the reliability of the systems. 

In distributed storage systems like P2P or RAID-based systems, erasure codes can improve the degree of reliability and the persistence of the data following the Information Dispersal Algorithm (IDA) proposed by Rabin \cite{rabin89}. This scheme is used is a large set of P2P systems proposals. Moreover, another benefit of erasure codes in this context is to decrease the average downloading time by increasing the data diversity in the network \cite{dairaine:peerfect}. 

Erasure codes have also applications in multimedia transmissions where the packet losses can not be recovered by classic retransmissions because of real-time constraints. Similarly, the reliability of multicast transmissions can be improved by codes since one repair packet can allow different receivers to recover different losses. 

Basically, an erasure code generates $n$ encoding symbols from a set of $k$ source symbols ($k\leq n$). The set of erasure codes can be split into two categories: MDS codes and non-MDS codes. The main property of MDS codes is that the $k$ source symbols can be recovered from any set of $k$ encoding symbols among the $n$ ones. For non-MDS codes, $(1+\epsilon)k$ symbols are needed on average to recover the $k$ source symbols, where $\epsilon > 0$. In counterpart, the encoding/decoding complexities of such non-MDS codes as Raptor \cite{raptor:Shokrollahi06} or LDPC codes \cite{rfc5170} is much lower than with MDS codes.

MDS codes can be built from Vandermonde matrices, like  Reed-Solomon codes \cite{fec:rizzo,rfc5510}, or from Cauchy matrices \cite{xor-Cauchy,pl:06:ocr,dairaine:peerfect}.  The optimal erasure recovery capability of MDS codes is necessary in many applications like distributed storage (peer-to-peer distributed storage or RAID-based systems) or multimedia/multicast transmissions. However, the quadratic encoding/decoding complexities of the codes mentioned restrain their practical use to applications needing short length codes  (in practical, up to $255$ symbols). 

   
The objective of our work is to design new MDS codes that support fast (sub-quadratic) operations. 

A recent work \cite{didier09} proposed a Reed-Solomon encoding/decoding algorithm based  on Walsh transforms with complexity $O(q\log^2 q)$, where $q=2^m$ is the size of the finite field. In order to go one step further towards sub-quadratic algorithms, we propose here  Reed-Solomon codes defined over a finite field $\mathbb{F}_p$, where $p$ is a Fermat prime, i. e. $p=2^{2^k}+1$. The interest of this choice is twofold: first, as noted in \cite{dairaine:peerfect}, the addition and multiplication operations in this finite field are performed very efficiently by modern computers since the processors are optimized for integer operations. The second (and the main) interest is that, unlike finite field of cardinal $2^m$, these finite fields support Fast Fourier Transform (FFT)-like algorithm with complexity $O(n\log n)$ allowing very fast polynomial evaluations/interpolations. This algorithm, called Fermat Number Transform (FNT), uses the \emph{divide and conquer} approach \cite{Cooley65} of FFT by replacing the $n^{th}$ roots of unity in the complex field by a $n^{th}$ root of unity in this finite field.

Even if the definition of Reed-Solomon codes in such field was already proposed \cite{1457487}, our work is, at our knowledge, the first complete description of the different steps of the encoding and decoding algorithms expressed in terms of FNT with an accurate analysis of the complexities. In particular, the use of recent results on polynomial multiplication \cite{HaQuZi04} allows us to reduce significantly the practical complexity (see Section \ref{sec:encodingdecodingScheme}). We show that the complexity of encoding and decoding are $O(k\log^2 k+n\log n)$ for the symbol erasure channel. Since the term $k\log^2 k$ only concerns the location of the erasure in a codeword, these operations are only done once per packet for a network transmission, or, once per file chunk for a peer-to-peer network. It follows that the effective complexity is $O(n\log n)$, with a low factor (see details in section \ref{sec:complexityAnalysis}). These results are confirmed by an analysis of the performance of our codec which is compared to other available MDS codecs (see Section \ref{sec:simulations}).

Another benefit of our approach is that the connection of erasure codes with FNT could allow to re-use the optimized software and hardware implementations \cite{ShuguoLi2009449} developed for the applications of FNT in various domains like multi-precision multiplication, audio or image filtering.

\section{FNT-based encoding/decoding scheme in a Fermat Field}
\label{sec:encodingdecodingScheme}

\subsection{The Fourier Transform over $GF(q)$}

Assuming that $q$ is prime, the values $0,1,...,q-1$ form a finite field where addition and product are processed modulo $q$.
This finite field is known as the Galois field $GF(q)$ \cite{1457487}.

In finite fields, the order of a number is defined as the lowest power of the number that equals $1$ modulo $q$.
An element of the field is called a \textit{primitive root} of the field if its order is $q-1$ \cite{MWSl77}.
For exemple, $3$ is a primitive root in $GF(65537)$, because $3^{65536} \equiv 1\ mod\ 65537$ and for each $0<i<65536$, $3^{i} \neq 1\ mod\ 65537$.

The principles of the Fourier transform can be extended to these finite fields, as introduced by Pollard \cite{Pollard:1971:FFT}. Let $r$ be an element of order $n-1$ in the field.
In this case, the discrete Fourier transform (DFT), which takes a vector $a=(a_0,...,a_{n-1})$ of size $n$ as input in $GF(q)$, returns:

$$
A_{j} = \sum_{i=0}^{n-1}{a_{i}r^{ij}}\ ,\ 0\leq j\leq n-1,\ n\leq q
$$

where $A$ is a vector of size $n$. In the same way, the inverse DFT (DFT$^{-1}$) can be defined as:

$$
a_{j} = \frac{1}{n}\times \sum_{i=0}^{n-1}{A_{i}r^{-ij}}\ ,\ 0\leq j\leq n-1,\ n<q.
$$

Fermat Number Transform is used in finite fields in order to process the DFT. FNT is the equivalent of the FFT algorithm on Fermat fields \cite{Cooley65}.
Thanks to a \emph{divide and conquer} approach, the complexity of the FNT of size $n$ and its inverse, FNT$^{-1}$, reduces the process of the DFT to $O(n\log n)$.

The FNT can also be represented by a $n\times n$ square matrix. This matrix is a special case of a Vandermonde matrix on special set $1,r,r^2,...r^{n-1}$, where $r$ is of order $n-1$. Since these values are pairwise distinct, the matrix is invertible. The first $k$ rows of this FNT matrix form the generator matrix of a Reed-Solomon code \cite{MWSl77}.

Another way to represent the FNT comes from polynomials. Indeed, if $a$ represents a vector of size $n$, we define $a(x)$ as the polynomial $\sum_{i=0}^{n-1}{a_{i}x^{i}}$. Thus, the FNT can be viewed as the evaluation of $a(x)$ on the points $1,r,r^2,...,r^{n-1}$, $A_j = a(r^j),\ 0\leq j\leq n-1$, which form a geometric sequence.

\subsection{Design of the code on $GF(65537)$}

In this section, we will present the FNT over $GF(65537)$ which allows to treat all codewords as 16-bit values, except 65536. This case can be treated easily, by sending the positions where this value is used in the encoding symbols header, if relevant.

As we have seen previously, $3$ is of order $2^{16}$. Thus, FNTs of size $n=2^i$ can be produced with the root $r=3^{2^{16-i}}$.

The objective is to encode $k$ source symbols to create $n$ encoding symbols, $k\leq n<65537$. As stated previously, the FNT matrix represents an MDS code for each couple $(n,k)$. For the sake of simplicity, we will take the hypothesis that $k$ is a power of 2, without losing generality. The case when $k$ is not a power of 2 can be treated as well, with padding on the FNTs that will be used.
In the same manner, we take also $n$ as a power of 2. The case where $n$ is not a power of 2, can be seen as a puncturing of the code.

Let $s = (s_0,s_1,...,s_{k-1})$ be the source vector of size $k$ and its associated polynomial $s(x)$. Let $e = (e_0,e_1,...,e_{n-1})$ be the encoded vector of size $n$.
The encoding step consists in simply applying the FNT on the vector $(s_0,s_1,...,s_{k-1},0,...,0)$ of size $n$. If $r$ represents the root of unity for the FNT of size $n$, we have also:

$$
  e_{i} = s(r^i),\ 0\leq i\leq n-1
$$

On the decoder side, some of the information in $e$ could have been lost. If not, the decoding is simply a product of $e$ by the FNT$^{-1}$ matrix. The decoded symbols are then the first $k$ symbols.
However, in most cases, at least one of the encoded symbols can have been lost. As we are dealing with an MDS code, decoding is possible if at least $k$ encoded symbols have been received. From a polynomial view, it means that at least $k$ points of evaluation have been received for a polynomial of degree $k-1$, which is a classical interpolation problem.
This field has been heavily studied for hundreds of years and many algorithms have emerged to solve this problem \cite{salvy}. Historically, Lagrange and Newton polynomial interpolation have been the most widespread solutions for this problem.
However, in our case, the main disavantage of these algorithms is their quadratic complexity. 

Let $(x_0,x_1,...,x_{k-1})$ be the first $k$ evaluation points received by the decoder. In fact, this vector represents a sub-sequence of the geometric sequence $1,r,r^2,...r^{n-1}$.
The decoder has then received the values $(s(x_0),s(x_1),...,s(x_{k-1}))$. Let this vector of size $k$ be called $v$. The Lagrange interpolating polynomial $P$ is then:

$$
P(x) = \sum_{i=0}^{k-1}{\left( v_{i}\times\prod_{0\leq j\leq k-1}^{j\neq i}{\frac{x-x_j}{x_i-x_j}}\right) }
$$

where the coefficients of $P$ are the coefficients of the source symbol $s$. Using the following definitions:

$$
A(x) = \prod_{j}{(x-x_j)}
$$

and

$$
A_i(x) = \prod_{j\neq i}{(x-x_j)} = \frac{A(x)}{x-x_i}
$$

we can see that for $i = 0...k-1$, $A(x_i) = 0$ and for each $j\neq i$, $A_i(x_j) = 0$, the Lagrange interpolating polynomial can also be rewritten as following:

$$
P(x) = \sum_{i=0}^{k-1}{v_i \frac{A_i(x)}{A_i(x_i)}} = A(x) \times \sum_{i=0}^{k-1}{\frac{v_i/A_i(x_i)}{x-x_i}}
$$

Let us define the derivative of the polynomial $A$:

$$
A^{'} = {(\prod_{j}{(x-x_j)})}^{'} = \sum_{i=0}^{k-1}{\prod_{j\neq i}{x-x_j}} = \sum_{i=0}^{k-1}{A_i}
$$

then one can remark that $A_i(x_j) = 0$ if $i\neq j$, meaning that for each $i = 0,1,...,k-1$ we have $A_i(x_i) = A^{'}(x_i)$.
This result is crucial as it means that we can use directly the derivative of $A$ in order to process all the $A_i(x_i)$ of the interpolating polynomial.

The algorithm for processing the interpolating polynomial, and then, recovering the source symbols is:

\begin{enumerate}
 \item Calculate the polynomial $A(x)$
 \item Derivate the polynomial to obtain $A^{'}(x)$
 \item Evaluate $A^{'}(x)$ on $(x_0,x_1,...,x_{k-1})$
 \item Process all the $v_i/A_i(x_i)$ in $n_i$
 \item Calculate $\sum_{i=0}^{k-1}{\frac{n_i}{x-x_i}}$
 \item As the denominator of the previous fraction is $A(x)$, the polynomial $P(x)$ is directly its numerator.
\end{enumerate}

We will detail further the complexity of each item. Most of these items depend heavily on polynomial products.
Hence, the design of an efficient method for processing the product of two polynomials will directly affect the complexity of the interpolation algorithm.
In the following development, we will introduce $M(n)$, which will correspond to the cost of multiplicating two polynomials of degree strictly lower than $n$.
Let us detail the complexity of $M(n)$. 

The classic polynomial product is unusable here because of its quadratic complexity. Using the fact that some products can be replaced by sums of already processed terms, Karatsuba \cite{KaOf63} has reduced the complexity bound to $O(n^{log_2 3})\simeq O(n^{1.59})$. This concept has been generalized by Toom and Cook. Toom-$k$ algorithms can lead to a complexity of $O(n^{\frac{log(2k-1)}{log(k)}})$. However, the constant factor hidden in the big-O notation, which fastly grows with $k$, prevents its practical use for $k>4$. Our approach, is based on the well-known Sch\"{o}nhage-Strassen algorithm \cite{SCHONHAGE71A}. This algorithm relies on FFTs to process the product of two polynomials. When all roots of unity are available, like in Fermat fields, the complexity of this algorithm is the complexity of processing 3 FFTs (FNTs in our case). It results that the complexity of polynomial multiplication is, in our case $M(n) = O(n\log n)$.\newline

In step 1, the product of $k$ polynomials of degree 1 has to be processed. The resulting polynomial $A(x)$ is then a polynomial of degree $k$.
Using a \textit{divide and conquer} method, the cost of step 1 is $O(M(k)\log k)$. Step 2 and step 4 have obviously a linear cost $O(n)$. 

In step 3, the problem is to evaluate a polynomial of degree $k-1$ on $k$ points. At this point, it may be noticed that the evaluation points form a sub-sequence of a sequence in geometric progression.
It does mean that if the evaluation of the polynomial $A^{'}$ is known on the geometric sequence of factor $r$, the knowledge on the evaluation points is then direct.

Let $A^{'}(x) = \sum_{i=0}^{k-1}{a_{i}^{'}x^{i}}$, $k<n$, and for $i=0,...,2n-2$, let $t_i=i(i-1)/2$, and the sequence $b_i=r^{t_i}$.
For $i=0,...,n-1$, let $c_i$ be defined by $c_i=a_{i}^{'}/b_i$. One may notice here that $b_{i+1}=q^ib_i$, and then we have:

$$
A^{'}(r^i) =  \sum_{j=0}^{n-1}{a_{j}^{'}r^{ij}} = \frac{1}{b_i}.\sum_{j=0}^{n-1}{a_{j}^{'}b_{i+j}}
$$

From this statement, we can see that the values of $A^{'}(r^i)$ can be viewed as the coefficients of the degrees $n-1,n,...,2n-2$ of the polynomial product of $\sum_{i=0}^{n-1}{a_{i}^{'}x^{n-i-1}}$ by $\sum_{i=0}^{2n-2}{b_ix^i}$.
Step 3 of the interpolation algorithm can be reduced to the product of two polynomials, and most importantly to the coefficients located in middle positions. The cost of step 3 is then $O(M(n))$ operations.

As the last step of the algorithm is trivial, the only remaining complexity to determine is step 5. In this step we have to determine the numerator of the following sum:

$$
\frac{P(x)}{A(x)} = \sum_{i=0}^{k-1}{\frac{n_i}{x-x_i}}\ mod\ x^n
$$

with the $n_i$ already determined in step 4. Using the Taylor series of $1/(x_i-x) = \sum_{j}{{x_i}^{-j-1}x^j}$ we can rewrite the sum as:

$$
\sum_{i=0}^{k-1}{\frac{n_i}{x-x_i}}\ mod\ x^n = -\sum_{i=0}^{k-1}{\left( \sum_{j=0}^{n-1}{n_i{x_i}^{-j-1}x^j}\right) }
$$

Then, we will also use the fact that the sequence $x_i$ is a sub-sequence of a sequence in geometric progression. We can then define each $x_i$ as a power of the root, $x_i=r^{z_i}$ with $z_i<n$. Then the sum becomes:

$$
\sum_{i=0}^{k-1}{\frac{n_i}{x-x_i}}\ mod\ x^n = -\sum_{j=0}^{n-1}{N^{'}(r^{-j-1})x^j}
$$

where $N^{'}(x)$ is the polynomial $\sum_{i=0}^{k-1}{n_ix^{z_i}}$. Since the points $1,r^{-1},r^{-2},...r^{-n}$ are in geometric progression, we can conclude that the complexity of step 5 is also $O(M(n))$.\newline

The decoding cost of the DFT-based code with this algorithm, in the general case, is $O(k\log^{2}k+n\log n)$ operations, with $k<n$.

It may be noticed that steps 1 to 3 do not depend on the received values but only on their positions. It means that if the received positions are known to be static, these steps have to be processed only once.
In the case of data transmission, data units can be UDP or RTP packets in the case of network transmission, or file chunks in the case of peer-to-peer networks. In the first case, the size of the packets can be up to 1500 bytes, and in the second case up to several megabytes.
This means that the positions are static for hundreds to millions symbols, which are two bytes long here, and thus, the complexity of these steps, mainly $O(k\log^2k)$ can be neglected compared to the other steps.

In conclusion, in practice, the complexity of the decoding algorithm falls to $O(n\log n)$.

\subsection{Toward a systematic code}

In network transmissions, it is often essential to have a systematic code, i.e. a code where the source data units are included in the encoded data units. Using the same algorithm, it is possible to design a systematic code.
Let $s=(s_0,s_1,...s_{k-1})$ be a source vector of size $k$. Interpolating these values with the previous algorithm on the first $k$ positions will lead to the ``decoding'' of $k$ intermediate symbols, $i=(i_0,i_1,...,i_{k-1})$.
Then producing the DFT of the vector $i$ with the same fast algorithm as the non-systematic case will produce the encoded vector $e$ where the first $k$ positions will be the systematic positions.

On the decoder side, if some source symbols (i.e. systematic positions) are missing, the same decoding as before on the first $k$ received positions is applied. This allows to recover the vector of intermediate symbols $i$. Then, the same DFT as the one used during encoding is applied to recover the full encoded vector $e$ whose first $k$ positions are the source symbols.

The practical complexity of a systematic version remains $O(n\log n)$, however with a greater constant term.

\section{Complexity analysis}
\label{sec:complexityAnalysis}

In this section, we will detail the complexity of the encoder and the decoder in both non-systematic and systematic coding, in terms of numbers of FNT of size $n$. The reason is that the big-O notation hides the constant term which has a major practical importance.
\footnote{It is also worth noticing that classic quadratic encoding and decoding schemes are also available for these codes.}

First of all, the complexity of the interpolation algorithm has to be discussed, and more precisely step 5, as it is in practice, the most costly step. In step 5, we first evaluate the polynomial $N^{'}$ on $n$ positions.
As we have seen before, this evaluation is equivalent to the multiplication of two polynomials of respectively degree $n-1$ and $2n-2$. As we stated in the beginning, $n$ is a power of two, meaning that the product is based on FNTs of size $4n$.
As seen in section \ref{sec:encodingdecodingScheme}, the only useful coefficients are the middle ones, so we can apply the Middle Product algorithm (MP) \cite{HaQuZi04}, which helps to reduce the FNTs size to $2n$. The latter element of step 5, is to multiply the resulting sum by $A(x)$. Since $k<n$, the product of the sum by $A(x)$, whose only the first $k$ coefficients are kept, is always of degree strictly greater than $n$ but lower than $2n$.
To summarize, the cost of interpolation, in terms of FNT$_n$, is 3FNT$_{2n}$+FNT$_{2n}$ = 4FNT$_{2n}$. As the FNT has a logarithmic cost, for reasonable sizes, we can say that FNT$_{2n} \simeq$ 2FNT$_n$, meaning that the interpolation cost is roughly 8FNT$_n$.\newline

For the encoder, we propose three algorithms,

\begin{itemize}
 \item The FNT-based algorithm described above for both systematic and non-systematic case;
 \item The direct encoding of the n-k non-systematic symbols by a quadratic complexity algorithm in the systematic case;
 \item The processing of the k intermediate symbols with a direct product and then a FNT, in the systematic case;
\end{itemize}

On the decoder side, we will use two algorithms,

\begin{itemize}
 \item The FNT-based interpolation algorithm described above for both systematic and non-systematic case;
 \item The direct decoding of the missing symbols;
\end{itemize}

The complexities are summarized in Table \ref{tab:encoding} and Table \ref{tab:decoding}.

\begin{table}[ht]
\begin{center}
\footnotesize{
\begin{tabular}{|r|c|c|}
\hline
    &  Non-Systematic Encoder &  Systematic Encoder \\
\hline
FNT-based	 &  1FNT$_n$ &  9FNT$_n$  \\
\hline
Direct Encoding      & -- &   O((n-k).k)  \\
\hline
Direct Interm. \& FNT  &  -- &  O(k$^2$)+FNT$_n$ \\
\hline
\end{tabular}}
\end{center}
\caption{Complexity of the encoder for both non-systematic and systematic case}
\label{tab:encoding}
\end{table}

\begin{table}[ht]
\begin{center}
\footnotesize{
\begin{tabular}{|r|c|c|}
\hline
    &  Non-Systematic Decoder &  Systematic Decoder \\
\hline
FNT-based	 &  8FNT$_n$ &  9FNT$_n$  \\
\hline
Direct Decoding      & -- &   $<$O((n-k).k)  \\
\hline
\end{tabular}}
\end{center}
\caption{Complexity of the decoder for both non-systematic and systematic case}
\label{tab:decoding}
\end{table}

For the systematic case, direct encoding and decoding methods are more adapted than FNT-based for small sizes. However, the last method for the encoder is well suited to small rate codes, i.e. k$<<$n.

\section{Simulation results}
\label{sec:simulations}

\begin{figure}[htb]
  \begin{center}
   \includegraphics[width=.485\textwidth]{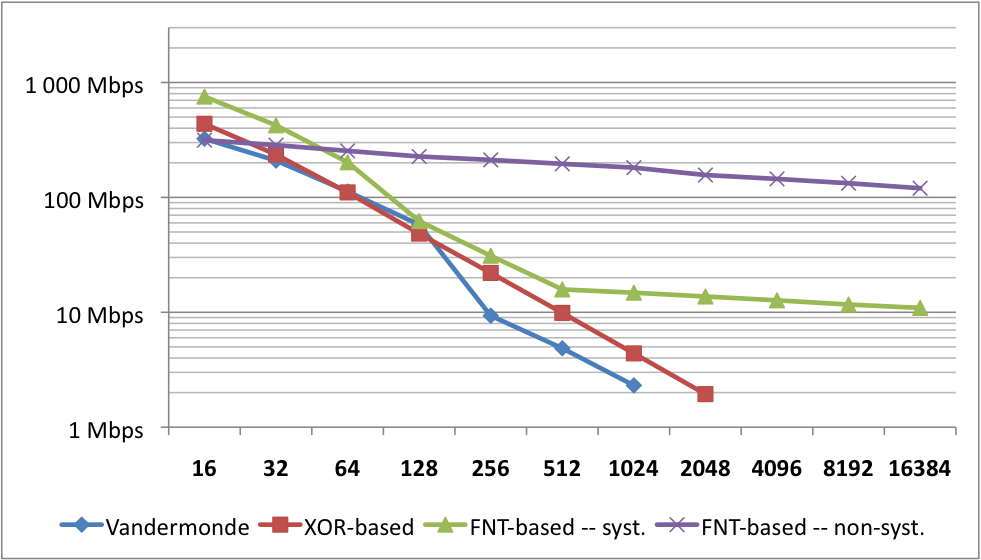}
    \end{center}
    \caption{Encoding speed for Vandermonde, XOR-based and FNT-based encoders for different source numbers}
   \label{fig:encode}
\end{figure}

\begin{figure}[htb]
  \begin{center}
   \includegraphics[width=.485\textwidth]{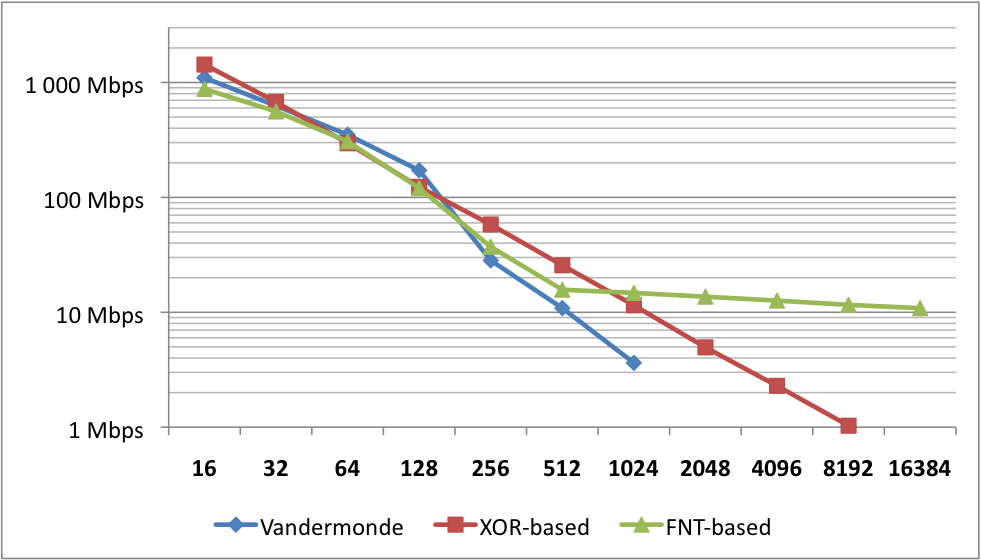}
    \end{center}
    \caption{Decoding speed for Vandermonde, XOR-based and FNT-based decoders for different source numbers}
   \label{fig:decode}
\end{figure}

We have implemented \cite{codefnt} a FNT-based algorithm, with the encoders and decoders described above. For k$\leq$256, quadratic complexity algorithms are used. Karatsuba and quadratic algorithms are used for polynomial products of degree less than 64.
We have tested also the Reed-Solomon implementation by Rizzo \cite{fec:rizzo}, named Vandermonde implementation, and the XOR-based implementation \cite{xor-Cauchy}, at their optimal tuning. It is important to notice that compared to the other RS algorithms presented here, FNT-based codes do not require any tuning on the field size.
The codes were tested on a Intel Core 2 Extreme@3.06Ghz on Mac OS X 10.5 with 64-bit compilation. Fig. \ref{fig:encode} provides the encoding speed for the three algorithms on a systematic case, and also the speed of non-systematic code for FNT-based algorithm, for a coderate 1/2.
Fig. \ref{fig:decode} shows the corresponding decoding speed. The speed of the non-systematic case is not represented in this figure, as it is slightly the same as that of the systematic case, as seen in Table \ref{tab:decoding}.

In the systematic scenario, FNT-based algorithms operates as fast as Vandermonde and XOR-based implementations for small sizes. The underlying algorithm is roughly equivalent to the other ones, but performs on integer computations, instead of binary polynomials.

However, as soon as the source size achieves hundreds of symbols, the logarithmic complexity algorithm of the FNT-based code allows a vast improvement compared to the other codes. In addition, the speed of the encoder/decoder decreases only slow, and speeds over 10Mbps are achieved with more than 10,000 source symbols.

For the non-systematic case, the encoder is about 10 times faster than the systematic one. This is in line with the complexity analysis.

\section{Conclusion and future work}

This paper presents the first implementation of an MDS code whose encoding and decoding complexity is $O(n\log n)$.
These results open the way for the practical use of MDS codes for sizes of thousands elements. Moreover, the speed of the encoder, which is roughly the speed of a FNT, allows many direct applications.
Among them, distributed data backup and peer-to-peer networks are areas where the encoding speed is critical, as well as the number of nodes implicated. With practical speeds of over 100Mbps, FNT-based coding is answering this issue, while offering a practically usable decoding speed. In deep space applications, the encoding cost could also benefit to low power systems.

As the speed of the algorithm heavily depends on the FNT, many ways can be explored for its optimization. Indeed, the FNT, which is a special case of FFT could benefit from parallel computing (multi-threading, GPU computing). Another interesting point is the re-use of existing optimized software and hardware implementations \cite{gmp}\cite{ShuguoLi2009449}.

\section*{Acknowledgements}

The authors would like to thank Vincent Roca and Mathieu Cunche for their useful comments.



%

\bibliographystyle{IEEEtran}
\bibliography{fftcodes}
%
%
%
%

\end{document}